\documentclass[10pt,twocolumn]{article}

\usepackage[margin=1in]{geometry}

\usepackage{booktabs}
\usepackage{times}
\usepackage{microtype}
\usepackage{graphicx}
\usepackage{amsmath, amssymb}
\usepackage{array}
\usepackage{listings}

\usepackage{xurl}

\usepackage{enumitem}
\setlist{nosep}

\usepackage[square, numbers]{natbib}
\usepackage{authblk}

\usepackage[hidelinks]{hyperref}
\hypersetup{
    colorlinks=true,
    linkcolor=black, 
    filecolor=black,
    citecolor=black,
    urlcolor=blue
}

\title{A Marketplace for AI-Generated Adult Content and Deepfakes}
\author[1,$\diamond$,$\dagger$]{Shalmoli Ghosh}
\author[2,$\diamond$]{Matthew R. DeVerna}
\author[1]{Filippo Menczer}

\affil[1]{\textit{Indiana University Bloomington}}
\affil[2]{\textit{Stanford University}}
\affil[$\diamond$]{\footnotesize These authors contributed equally.}
\affil[$\dagger$]{\footnotesize To whom correspondence should be addressed.}
\affil[ ]{\texttt{shaghosh@iu.edu}, \texttt{mdeverna@stanford.edu}, \texttt{fil@iu.edu}}
\date{}  

\begin{document}
\maketitle

\begin{abstract}
    Generative AI systems increasingly enable the production of highly realistic synthetic media.
    Civitai, a popular community-driven platform for AI-generated content, operates a monetized feature called \textit{Bounties}, which allows users to commission the generation of content in exchange for payment.
    To examine how this mechanism is used and what content it incentivizes, we conduct a longitudinal analysis of all publicly available bounty requests collected over a 14-month period following the platform's launch.
    We find that the bounty marketplace is dominated by 
    tools that let users steer AI models toward content they were not trained to generate. 
    At the same time, requests for content that is ``Not Safe For Work'' are widespread and have increased steadily over time, now comprising a majority of all bounties. 
    Participation in bounty creation is uneven, with 20\% of requesters accounting for roughly half of requests. Requests for ``deepfake''---media depicting identifiable real individuals---exhibit a higher concentration than other types of bounties.
    A nontrivial subset of these requests involves explicit deepfakes despite platform policies prohibiting such content.
    These bounties disproportionately target female celebrities, revealing a pronounced gender asymmetry in social harm.
    Together, these findings show how monetized, community-driven generative AI platforms can produce gendered harms, raising questions about consent, governance, and enforcement.
\end{abstract}

\section{Introduction}
\label{sec:intro}

Text-to-image generative AI has fundamentally transformed image creation, enabling anyone with internet access to produce highly realistic images from simple text descriptions.
Modern diffusion models such as Stable Diffusion~\cite{podell2023sdxl}, DALL-E~\cite{ramesh2021zero}, and Midjourney allow users to create everything from artistic illustrations and product mockups to photorealistic portraits and scenes that are increasingly difficult to distinguish from authentic photographs~\cite{corvi2023detection, cazenavette2024fakeinversion}.

The rapid adoption of these models has democratized visual content generation, giving rise to a new class of online platforms that blend social interaction with AI-powered content creation. 
Platforms such as Civitai,\footnote{\href{https://civitai.com}{civitai.com}} PixAI,\footnote{\href{https://pixai.art/en}{pixai.art/en}} and TensorArt\footnote{\href{https://tensor.art}{tensor.art}} have become popular hubs where users gather around shared tools, models, and generated outputs.
Users in these new environments simultaneously act not only as creators and consumers of AI-generated media, but also of the underlying generative tools themselves.
These platforms therefore constitute a distinct digital space, raising new questions about their governance and content moderation, as well as the social implications of widespread access to powerful generative technologies~\cite{ThielGenerativeML2024}.

A growing body of research highlights how these platforms can also facilitate the production of harmful or misleading content.
Generative AI has been used to create political disinformation, identity fraud, child sexual abuse material, and non-consensual intimate imagery, often at low cost and high scale~\cite{fu2023misusing, Wei2024CivitaiOct, AI-faces, hawkins2025deepfakes, ciardha2025ai}. 
The ease with which such content can be generated and circulated exacerbates longstanding challenges around consent, authenticity, and accountability, motivating closer empirical scrutiny of how AI-centric platforms structure participation and enforce safeguards~\cite{ThielGenerativeML2024}. 

Against this broader landscape, the Civitai platform provides a particularly useful case for studying governance and incentives within these social generative AI ecosystems.
In the United States, Civitai has rapidly grown into the largest AI-centric content platform, reaching approximately 3 million registered users and 12–13 million unique monthly visitors by November 2024.
Founded in 2022, the platform has raised over US\$5 million in venture funding, including investment from Andreessen Horowitz, signaling increasing mainstream interest in these platforms~\cite{Perez2023}.
At the same time, Civitai's open structure has drawn criticism for enabling intellectual property violations~\cite{Nikken2023Nov} and facilitating the circulation of child sexual abuse material and non-consensual deepfakes~\cite{ThielGenerativeML2024, ThielCrossPlatform2023, Epstein2023Jun, Maiberg2023CSAMDec, hawkins2025deepfakes}.
These concerns echo earlier warnings about the societal effects of generative media~\cite{9123958, al2023impact, hancock2021social} and align with mounting evidence of increasing online harassment and gendered harm linked to deepfake technologies~\cite{amin2025women, lazard2025deepfake, akter2025emergence}. 

\begin{figure*}[t]
\centering
\begin{tabular}{cc}
\raisebox{2.12in}{(a)} &
\includegraphics[width=0.75\textwidth]{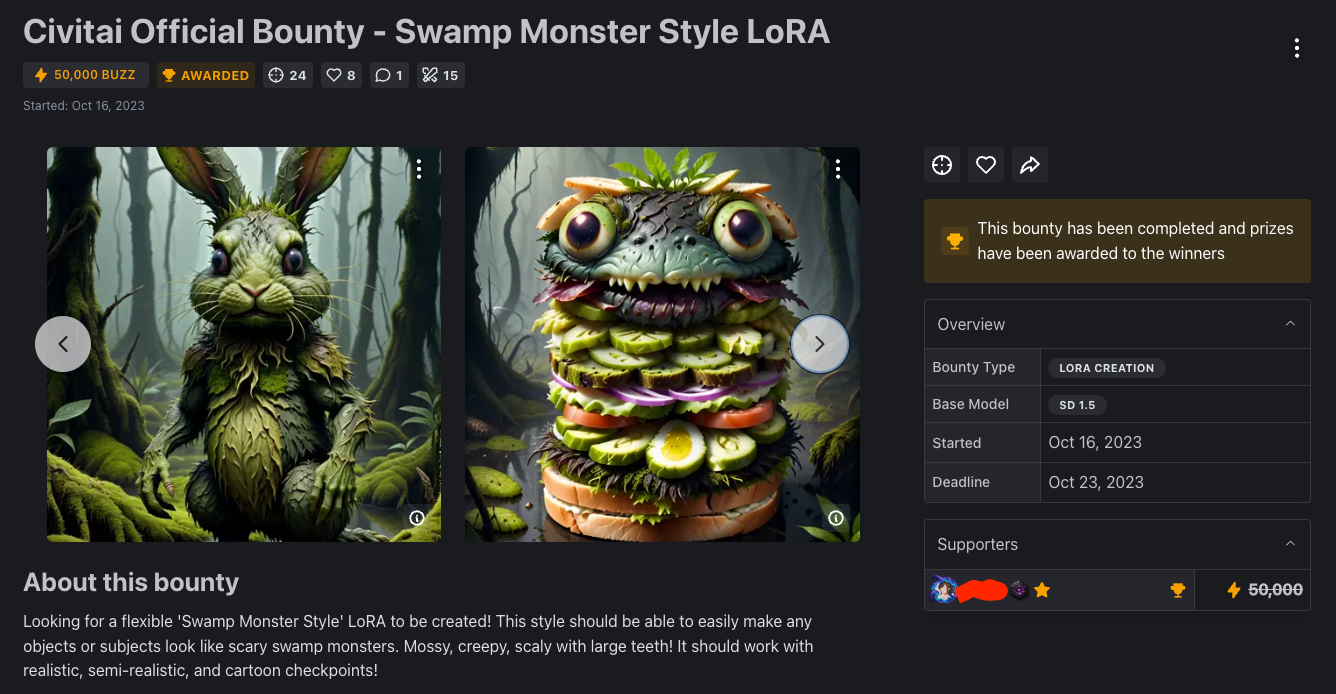} \\
\raisebox{2in}{(b)} &
\includegraphics[width=0.75\textwidth]{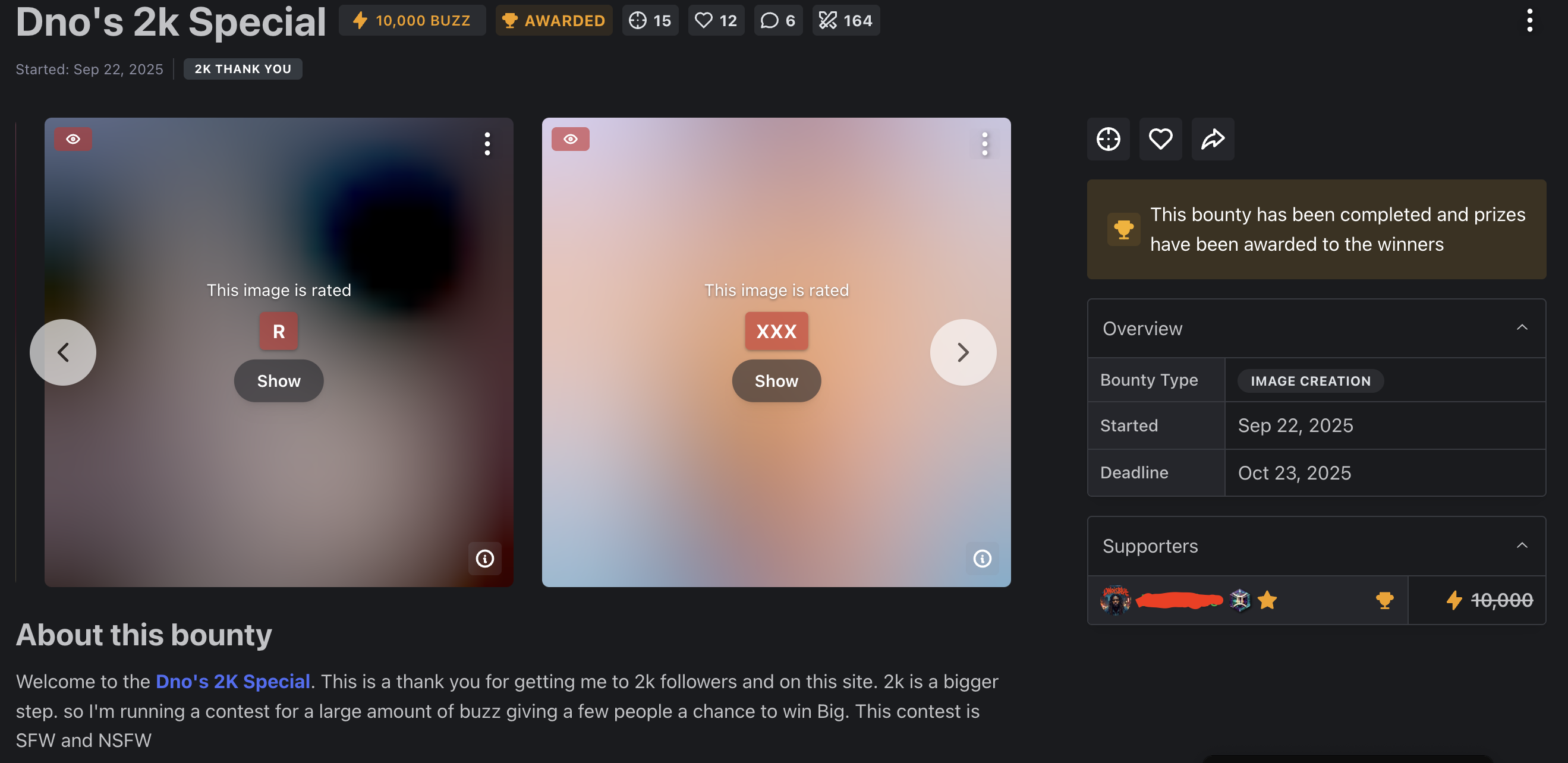} \\
\end{tabular}
\caption{Example Civitai bounties. (a)~A bounty with non-blurred images. (b)~A bounty having blurred images of different ratings. The bounty poster's names (under `Supporters') are hidden for privacy reasons.}
\label{fig:fig1}
\end{figure*}

Civitai differs from other generative AI platforms not only in scale, but also in platform design. 
While platforms such as PixAI and TensorArt primarily center on sharing, remixing, and showcasing AI-generated content, Civitai uniquely incorporates a formal bounty system that allows users to post paid requests for specific types of AI-generated media~\cite{Maiberg2023Nov}.
Figure~\ref{fig:fig1} presents two example bounties, including their titles, descriptions, and reference images.
Each bounty is posted by a user (referred to as a \textit{Supporter}) who offers virtual currency (\textit{buzz}) to reward one or more winning submissions.
Buzz can be purchased (1,000 buzz per US dollar) and spent within the platform.
Bounties progress through several states---Active, Awarded, or Expired---and user engagement is reflected through the number of likes, comments, and submitted entries.
Through this system, users who lack the time or technical expertise to generate desired content can offer monetary rewards for specific requests, creating incentives for other users to compete to produce and submit entries.
This feature introduces a structured marketplace for AI-generated content that directly coordinates demand, labor, and reward---an incentive mechanism that is absent from comparable platforms.

Empirical studies of Civitai have begun to characterize the content produced on the platform. 
\citet{Wei2024CivitaiOct} describe Civitai as an AI-generated content platform similar to social media, where users share generative models, showcase AI-generated images, and participate in community discussions. 
They conduct the first large-scale empirical study of the platform, spanning 87K models and 2M images available on the platform at the time of their research.
They analyze the themes, prompts, and engagement patterns of abusive images and models, finding that creators of such content often gain increased prominence.
\citet{palmini2024civiverse} introduce a dataset of 6M Civitai prompts and millions of images generated with these prompts, along with related metadata. 
Their analysis of features requested and excluded in prompts reveals a strong user preference for explicit content and raises concerns about the reproduction of harmful stereotypes.
Yet these studies focus primarily on generated images and models themselves, leaving the dynamics of Civitai's bounty marketplace largely unexplored.

Here, we address this gap by conducting the first systematic analysis of Civitai's bounty system, examining participation patterns and content incentives within the marketplace. 
Specifically, we want to characterize what content is allowed within bounty requests, capturing marketplace-level permissiveness rather than personal intent or final output usage.
Our research explores the following questions:
\begin{itemize}
    \item \textbf{RQ1:} What content types are requested in Civitai's bounty marketplace, and how prevalent are SFW (Safe For Work) versus NSFW (Not Safe For Work) requests?
    \item \textbf{RQ2:} How has the prevalence of SFW and NSFW bounty requests changed over time?
    \item \textbf{RQ3:} To what extent is the bounty system used for deepfake requests, and who is most frequently targeted?
    \item \textbf{RQ4:} How concentrated is bounty creation across users, and does this concentration differ for SFW, NSFW, and deepfake bounties?
    \item \textbf{RQ5:} How consistently are Civitai's deepfake governance mechanisms applied across bounty requests?
\end{itemize}

To answer these research questions, we conduct a large-scale, longitudinal analysis of all publicly available bounty requests from the platform's first 14 months, resulting in 4,847 bounties spanning 15 October 2023 to 30 January 2025.
We find that the proportion of NSFW bounty requests increased steadily over time, with nearly half of all bounties explicitly expressing an intent to generate NSFW content.
Deepfake requests, many of which are also NSFW, constitute a non-negligible share of the marketplace and disproportionately target female celebrities.
Our analysis further reveals strong concentration in participation, with a small group of users driving most deepfake-related activity.
Finally, we examine the platform's interventions for flagging deepfake content and assess their effectiveness across SFW and NSFW categories, identifying gaps in policy enforcement that warrant attention from platform moderators and regulators.

The remainder of the paper is organized as follows. 
Section~\ref{sec:methods} describes the data collection process and the content classification methods.
We describe the results in Section~\ref{sec:results}. 
Section~\ref{sec:disc} discusses implications, limitations, and future work.

\section{Data and methods}
\label{sec:methods}

To construct a comprehensive dataset of Civitai bounties,\footnote{\href{https://civitai.com/bounties}{civitai.com/bounties}}, we built a custom data-collection pipeline. Our code and data are publicly available.\footnote{\href{https://github.com/osome-iu/civitai_bounties_osome}{github.com/osome-iu/civitai\_bounties\_osome}}

We automated browser sessions and saved the complete HTML for each bounty page using the Python Selenium package.\footnote{\href{https://selenium-python.readthedocs.io}{selenium-python.readthedocs.io}} 
Civitai pages can render in two modes depending on a user's browsing preferences: default and ``blur mature content'' (Figure~\ref{fig:fig1}). 
Both rendering modes expose complementary information useful to our analysis.
In default mode, all images are visible and their URLs are included in the pages.  
When the ``blur mature content'' option is enabled, explicit images are blurred and their URLs are unavailable; instead, the page presents platform-provided explicitness ratings for each image, which may take labels such as ``R,'' ``X,'' or ``XXX'' (see Table~\ref{tab:nsfw-ratings} for all labels and their definitions; in rare cases, images are completely hidden with a warning about mature content). 
Therefore our data collection proceeds in two passes: a blurred pass to capture explicitness labels and all other bounty data (4,959 bounties), followed by an unblurred pass to capture image URLs (4,986 bounties). 
After both passes are complete, we retain only bounties for which we successfully collected both components.

\begin{table}
    \centering
    \caption{Civitai image content categories.}
    \begin{tabular}{>{\centering\arraybackslash}p{2cm}|>{\raggedright\arraybackslash}p{5cm}}
    \toprule
       Classification & Content \\ \midrule
        PG & Safe For Work. No naughty content!\\
        PG-13 & Revealing clothing/Sexy attire, Violence, Light gore\\
        R & Adult themes and situations, Partial nudity, Graphic violence and death\\
        X & Graphic nudity, Adult objects and settings\\
        XXX & Overtly sexual or disturbing graphic content\\ \bottomrule
    \end{tabular} \\
    \footnotesize\textit{Note:} See \href{https://education.civitai.com/civitais-guide-to-content-levels}{education.civitai.com/civitais-guide-to-content-levels} for more details.
    \label{tab:nsfw-ratings}
\end{table}

We run this pipeline over all publicly listed bounties available at the time of collection, yielding a total of 4{,}847 bounties spanning from October 15, 2023 to January 30, 2025. 
Of these bounties, 84.1\% were awarded, 14.4\% expired, and 1.5\% active at collection time. 
For each bounty, we extract the title, a description of the request posted by the supporter, timestamps (posting date and deadline), engagement metrics (counts of entries, comments, likes, and users tracking the bounty), and bounty type. 
Civitai supports seven bounty types reflecting the form of the requested output: \textit{models} (fully trained generative models), \textit{LoRAs} (Low-Rank Adaptations, fine-tuned versions of existing models that extend their capabilities), \textit{images} and \textit{videos} (generated media), \textit{datasets} (collections of training data), \textit{embeddings} (numerical representations for downstream use), and an \textit{other} category. 
We also record data related to all example images, which are uploaded by the supporter to illustrate how they envision the request. 
For each example image, we collect its URL and any platform-provided explicitness label.

\subsection{Labeling bounty themes}
\label{sec:labeling_themes}

A central goal of this study is to characterize the types of content requested and permitted in the bounty marketplace, including the extent to which it is used to solicit mature and deepfake content (\textbf{RQ1–RQ3}).
We therefore classify bounties as either SFW or NSFW using the platform's explicitness ratings. 
Following Civitai's moderation policy, any image labeled ``R,'' ``X,'' or ``XXX'' may contain ``adult themes'' or ``partial nudity'' and is therefore treated as NSFW (Table~\ref{tab:nsfw-ratings}). 
Because these ratings apply at the image level, we assign each bounty the most explicit rating present among its example images. 
Each bounty is thus labeled NSFW if any included image is rated ``R'' or more explicit. 
This approach reflects the supporter's intent: if even one example image is ``XXX,'' the request permits highly explicit content, regardless of the other images. 
Out of 4{,}847 bounties, 43\% are classified as NSFW and 17\% have at least one image labeled ``XXX.''

To contextualize how Civitai's platform-based explicitness ratings compare to an external automated moderation system, we also applied OpenAI's content moderation endpoint\footnote{\href{https://platform.openai.com/docs/guides/moderation}{platform.openai.com/docs/guides/moderation}} to all unblurred bounty images.
We pass to this endpoint image URLs under the default API settings.
For each image, the API returns category-specific scores indicating the presence of potentially harmful content (e.g., sexual, violent, or graphic material). 
Each category also includes a boolean \texttt{flagged} indicator, which is set to \texttt{True} when the model determines that the image violates the corresponding content category.

Among the 4,704 bounties with at least one image, 1,476 had at least one image flagged for one or more of four categories: sexual (1,092 bounties), violence (697), violence/graphic (36), and self-harm (23). 
We labeled a bounty as NSFW if any associated image was flagged by the API; otherwise it was labeled SFW.

We then compared these externally generated flags to Civitai's platform-provided labels at the bounty level to characterize differences in how explicit content is identified across moderation systems, computing Cohen's $\kappa = 0.52$, Matthews correlation coefficient $\text{MCC} = 0.53$, and $\text{accuracy} = 0.77$ (using Civitai labels as ground truth).  
If bounties whose most explicit rating is ``R'' are treated as SFW, agreement improves: $\kappa = 0.63$, $\text{MCC} = 0.63$, $\text{accuracy} = 0.84$. 
Overall, the comparison suggests that Civitai adopts a more conservative approach to content classification and presentation than OpenAI's moderation system, likely reflecting the challenges of governing a platform where highly graphic content is common. 
Because our analysis centers on content visibility, incentives, and moderation dynamics as they operate on Civitai itself, we rely on the platform's own explicitness labels in all subsequent analyses.

Leveraging only image content may underestimate the true volume of explicit bounty requests.
Users may request explicit content within the bounty title and/or description. 
For example, a poster may use a bounty request to generate NSFW content that they currently are unable to create by describing it with text.
To ensure we capture these cases, we also wish to classify bounties based on their central themes. 

After the image-based classification described above, we employed a large language model to assign a primary thematic category to each SFW bounty.
To select an appropriate model for this task, we first drew a sample of SFW bounties and applied manual annotations.  
We identified the three most prevalent bounty types that together account for approximately 95\% of all bounties: LoRA, image, and model (see Figure~\ref{fig:fig2}).
We then randomly sampled 50 bounties from each of these three types, yielding a total sample of 150 bounties. 

A human annotator conducted thematic analysis on the selected bounties.
While human annotation ensures contextual judgment in borderline cases, the choice of a single annotator was due to the substantial proportion of highly graphic or disturbing material on Civitai and the need to minimize exposure to this content. 
The procedure required the annotator to review the complete bounty content, including titles, descriptions, requirements, and any associated visual materials.
For each bounty, the annotator assigned a primary thematic category that best represented the central focus or purpose of the request. 
The resulting thematic taxonomy comprises eight categories: human deepfake; NSFW; human attributes; fictional characters; scenes, objects, and clothing; style and culture; miscellaneous; and open requests.
Table~\ref{tab:taxonomy} provides the definition of each theme. 

\begin{table*}[t]
\centering
\caption{Thematic taxonomy of bounty categories. Each bounty is assigned one of these categories. For overlapping labels, human deepfake has the highest priority and NSFW is the next. All other themes have the same priority.}
\label{tab:taxonomy}
\begin{tabular}{p{0.2\textwidth}p{0.75\textwidth}}
\toprule
\textbf{Theme} & \textbf{Definition} \\
\midrule
Human deepfake & Real person image generation requests \\
\addlinespace
NSFW & Adult-oriented material (either bounties having `R', `X', `XXX' rated images or having explicit mention of NSFW generation in the title and/or description of the bounty) \\
\addlinespace
Human attributes & Requests focused on specific physical or behavioral human characteristics \\
\addlinespace
Fictional characters & Virtual characters, fictional content, and entertainment media-based requests \\
\addlinespace
Scenes, objects, and clothing & Scenery, decorative objects, and apparel-related bounties \\
\addlinespace
Style and culture & Artistic style preferences and cultural aesthetic requests \\
\addlinespace
Miscellaneous & Uncategorizable content such as mascot generation, logos, t-shirt designs, model improvements, and watermark removal. The requests look for specific content generation. \\
\addlinespace
Open requests & Bounties where requesters asked for general content generation using specific models, contests, or broad creative challenges. The requests do not look for specific content generation. \\
\bottomrule
\end{tabular}
\end{table*}

Following human annotation, automated thematic classification of the bounties in the sample was conducted using two large language models: GPT-4o and GPT-4.1. 
Both models were tasked with extracting the best theme (see Table~\ref{tab:taxonomy}) for a given bounty. 
Both models received a detailed prompt that described the Civitai context, defined the eight thematic categories with detailed descriptions, specified explicit classification instructions, and provided the bounty title, description, and associated image URLs (see Appendix~\ref{app:B}). 
The prompt specified hierarchical prioritization rules: models should first identify potential human deepfake content; otherwise assess for NSFW material; and if neither applies, select one of the remaining themes. 
This hierarchy ensures that deepfake requests are identified regardless of explicitness and that NSFW requests are isolated prior to further thematic differentiation.
Models were instructed to analyze bounty titles, descriptions, and associated images to make thematic assignments.
The exact prompt is provided in the Appendix. 

To select the best model, we evaluated performance through comparison with human annotations on the sample bounties using the Krippendorff's $\alpha$ inter-rater reliability metric, yielding scores of $\alpha=0.7$ for GPT-4o and $\alpha=0.8$ for GPT-4.1. 
Measuring performance with aggregate F1 metrics also returned better performance for GPT-4.1 (macro-F1 = 0.75; weighted-F1 = 0.84) compared to GPT-4o (macro-F1 = 0.65; weighted-F1 = 0.77).
See Appendix~\ref{app:f1} for details. 
Based on its higher agreement with human annotations, GPT-4.1 was selected for automated thematic classification and subsequently employed for large-scale automated theme extraction across all 2,752 SFW bounties. 
We use the same prompt (see Appendix~\ref{app:B}).
Note that this automated approach could not be applied to bounties previously labeled as NSFW based on platform-provided ratings, because general-purpose GPT models consistently refuse to process prompts containing extremely explicit NSFW images, which are present in these bounties. 

\subsection{Labeling deepfake bounties}

The GPT 4.1 model labeled 315 bounty requests as `human deepfake.' 
Because these requests appeared to involve real individuals, a human annotator manually reviewed all identified cases to verify that each request targeted a real person. 
Sixteen bounties were found not to meet this criterion and were reclassified into the appropriate non-deepfake thematic categories.

In parallel, we sought to identify deepfake requests among bounties classified as NSFW based on platform-provided image ratings.
Given the GPT-model guardrails described earlier, we submitted to GPT-4.1 only the titles and descriptions of these NSFW bounties to identify potential deepfake requests.
The prompt we used is given in Appendix~\ref{app:C}. 
This process yielded 31 potential NSFW deepfakes. 
Manual review was again carried out by a single human annotator to minimize exposure to graphic content. 
Seven bounties were excluded following manual review for not targeting real individuals.

In total we identify and confirm 323 unique bounties that solicit human deepfake content within our dataset. 

\subsection{Concentration of bounty participation}

In addition to characterizing what types of content are requested on Civitai, a central question for platform governance concerns who drives these requests.
In particular, RQ4 asks whether participation in the bounty marketplace is broadly distributed across users or concentrated among a small subset of highly active requesters, and whether such patterns differ across SFW, NSFW, and deepfake content categories.

To address this question, we analyze the concentration of bounty creation across users.
Following thematic classification, we collapse bounties into three analytically relevant categories.
\textit{NSFW} includes all bounties classified as Not Safe For Work.
\textit{SFW} includes all remaining bounties classified as Safe For Work.
\textit{Deepfake} includes all bounties that request the generation of real individuals, regardless of whether they are classified as SFW or NSFW.
These categories are therefore not exclusive; deepfake bounties can fall within either the SFW or NSFW categories. 

For each category, we compute the number of bounties posted by each user and assess the resulting distribution using the Lorenz curve and Gini coefficient to quantify concentration.
The Lorenz curve plots the cumulative share of bounties against the cumulative share of users, ordered from most to least active.
The Gini coefficient summarizes this distribution as a single value between 0 and 1. 
Formally, let $x_i$ denote the number of bounties posted by user $i$, for $i = 1, \dots, n$. 
The Gini coefficient $G$ is defined as:
\[
G = \frac{\sum_{i=1}^{n} \sum_{j=1}^{n} |x_i - x_j|}{2n \sum_{i=1}^{n} x_i}.
\] 
$G=0$ indicates perfectly equal participation (all users post the same number of bounties) and $G=1$ indicates maximal concentration (a single user posts all bounties). 

\subsection{Deepfake intervention effectiveness}

Finally, RQ5 examines one of Civitai's approaches to moderating requests for deepfake bounties.
Specifically, the platform displays an informational notice on some bounty pages that request the reproduction of a real person's likeness.
This notice states that, out of respect for the individual and in accordance with platform rules, only work-safe images and non-commercial use are permitted, and it provides a mechanism for the individual or their legal representative to request removal of the bounty.

While the platform does not publicly document how this marker is assigned, its presence suggests that either automated detection mechanisms or supporter-provided disclosures are used to flag bounties involving real individuals.
To assess how this intervention is applied in practice, we examine its prevalence across bounties identified as deepfakes.

Specifically, for all bounties classified as human deepfakes---both SFW and NSFW---we record whether the platform-displayed notice is present on the bounty page at the time of data collection.
We then compute the proportion of deepfake bounties in each category that carry this marker.
This analysis provides a descriptive measure of how consistently the platform applies its existing intervention to deepfake-related requests and allows for comparison of intervention coverage across SFW and NSFW deepfake bounties.

\section{Results}
\label{sec:results}
\subsection{Bounty themes}

\begin{figure*}
\centering
\includegraphics[width=\linewidth]{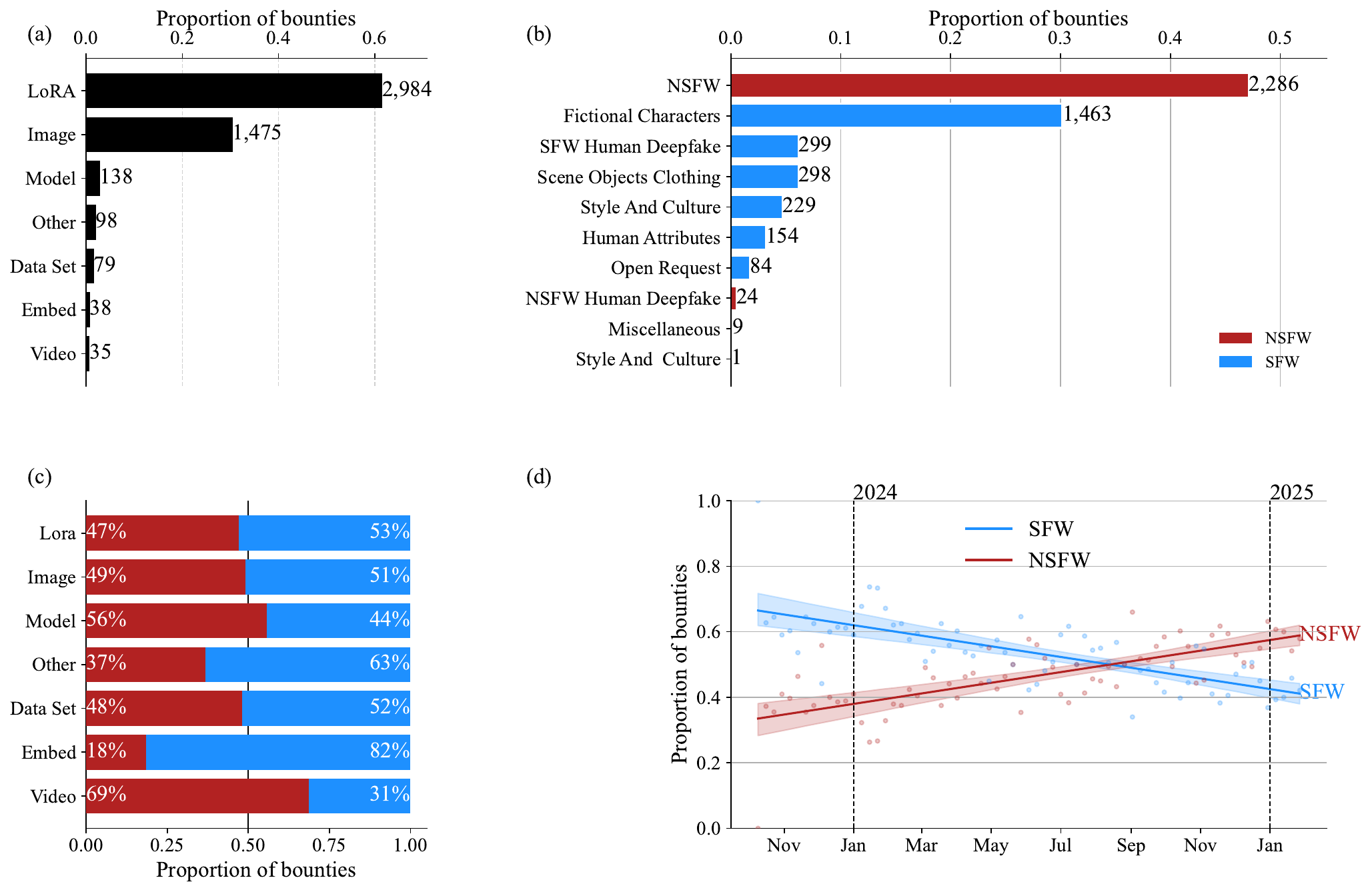}
\caption{Civitai's bounty marketplace shows a substantial and growing proportion of NSFW requests.
(a)~Distribution of bounty request types. 
(b)~Distribution of bounty themes.
(c)~Proportions of SFW (blue) and NSFW (red) bounties across different content types, highlighting that certain categories, videos in particular, have a higher proportion of NSFW requests. 
(d)~Proportions of SFW (blue) and NSFW (red) bounties over time, showing a decline in SFW and a rise in NSFW bounties. Each point represents a weekly proportion, with ordinary least squares regression fit lines and bootstrapped 95\% confidence intervals ($n=1{,}000$ samples).}
\label{fig:fig2}
\end{figure*}

We begin by characterizing the overall composition of the bounty marketplace (\textbf{RQ1}).
Figure~\ref{fig:fig2}(a) shows the distribution of bounty types.
The most common type is LoRA ($n = 2{,}984$), followed by requests for images ($n = 1{,}475$); together, these two categories represent 92\% of bounties).
Figure~\ref{fig:fig2}(b) presents the results of the thematic analysis.  
NSFW bounty requests dominate all other themes, accounting for 48\% of bounties when NSFW human deepfakes are included. 
The next most frequent theme is fictional characters (30\%).

To examine how mature content is distributed across bounty types and how it varies over time, we group the NSFW human deepfake themed bounties into the NSFW category.
All remaining themes (`SFW Human deepfake,' `Human attributes,' `Fictional characters,' `Scenes, objects, and clothing,' `Style and culture,' `Miscellaneous,' and `Open requests') are grouped within the SFW category.
Figure~\ref{fig:fig2}(c) reveals that a substantial portion of bounties request NSFW content across most bounty types: 69\% of video, 56\% of model, 49\% of image, and 47\% of LoRA bounties fall into this category.

We further analyze temporal trends by calculating the weekly proportion of NSFW bounty requests (\textbf{RQ2}).
As shown in Figure~\ref{fig:fig2}(d), the share of NSFW bounties increased steadily over time, surpassing half of all weekly bounty requests by September~2024 and continuing to rise in the subsequent months.
Our analysis further reveals that NSFW bounties paid on average \$6.00 (SD \$10.32) and facilitated roughly \$13.8K in total transfers. SFW bounties paid on average \$5.30 (SD \$10.12), facilitating roughly \$13.5K in total transfers. These figures should be considered noisy lower bounds.
Bounty comments suggest requesters can pay multiple entrants via unofficial means, leaving no structured metadata capturing those payments. 

\subsection{Deepfakes in bounties}

\begin{figure*}
\centering
\includegraphics[width=\linewidth]{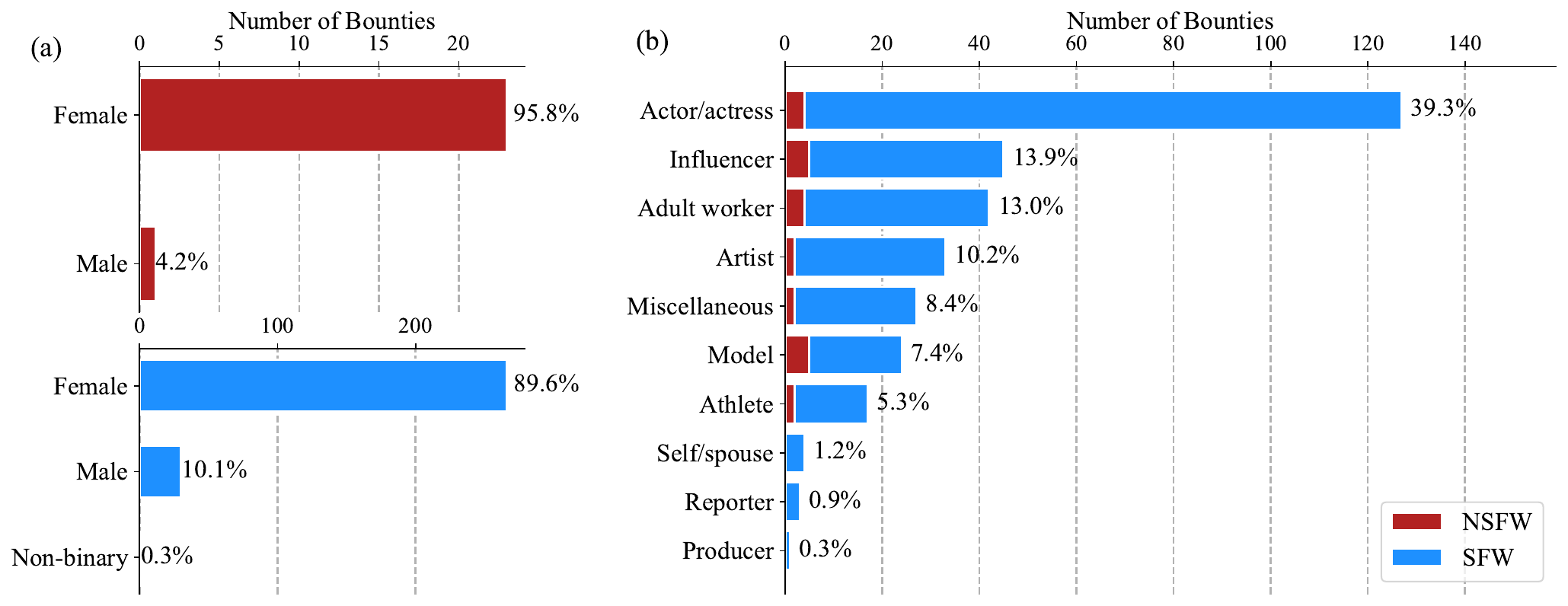}
\caption{Deepfake details among Civitai bounties. (a)~Distribution of targeted genders in the deepfake requests. (b)~Distribution of professions as found in public records (e.g., Wikipedia or social media profiles) related to the real persons targeted in the deepfake bounty requests. The ``self/spouse'' tag refers to bounties claiming to target the supporter or their spouse.}
\label{fig:fig4}
\end{figure*}

We next examine the prevalence and characteristics of deepfake requests in the bounty marketplace (\textbf{RQ3}).
As described earlier, we identify 323 unique bounties that solicit the generation of human deepfakes, of which 299 are classified as SFW and 24 as NSFW.

Figure~\ref{fig:fig4}(a) illustrates that there is a pronounced gender imbalance in the targets of these requests.
Deepfake bounties targeting women constitute the overwhelming vast majority, with 289 cases (266 SFW and 23 NSFW), compared to 31 targeting men (30 SFW and one NSFW), and a single SFW bounty targeting a non-binary individual. 
Two additional SFW bounties target multiple or unspecified individuals and are not assigned a single gender.

Figure~\ref{fig:fig4}(b) provides additional context by showing the professional backgrounds of targeted individuals.
Actors and actresses comprise by far the largest group, followed by smaller numbers of influencers, adult workers, artists, and models. 

\subsection{User concentration}

\begin{figure}
\centering
\includegraphics[width=.6\linewidth]{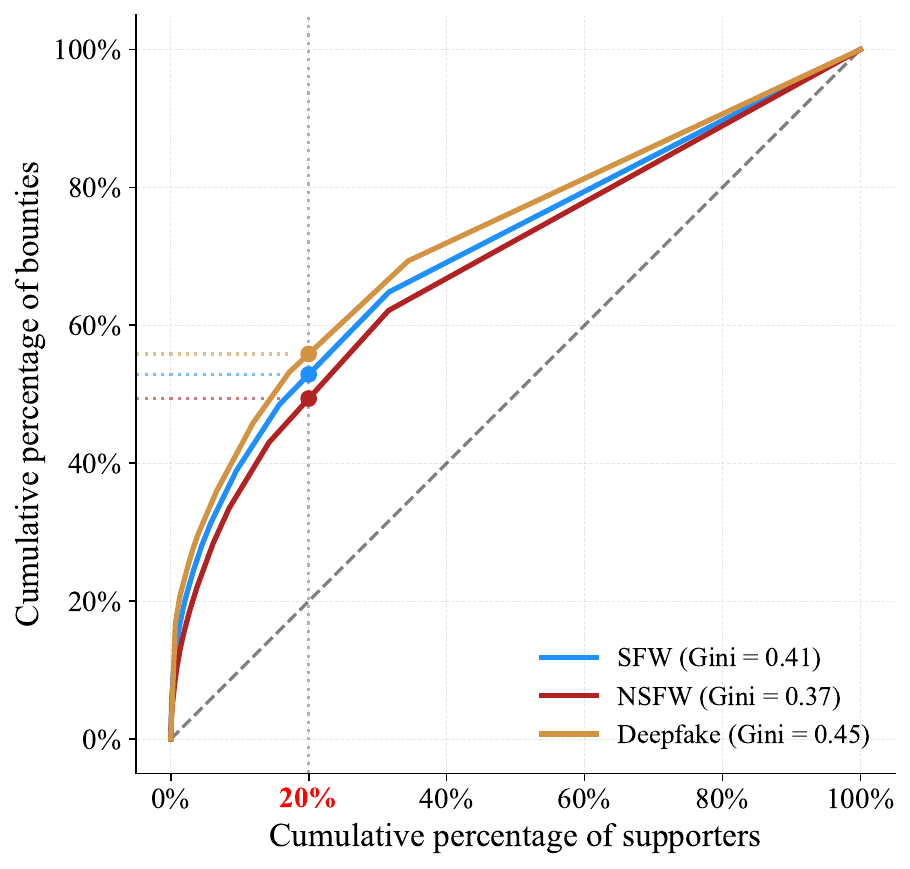}
\caption{Lorenz curves and Gini coefficients showing user concentration across bounty categories. The dashed line denotes perfect equality; greater distance above this line indicates that fewer users account for a larger share of bounty requests.}
\label{fig:fig5}
\end{figure}

We next examine how participation in the bounty marketplace is distributed across users (\textbf{RQ4}).
Figure~\ref{fig:fig5} shows Lorenz curves and corresponding Gini coefficients for SFW, NSFW, and deepfake bounties.

Across all three categories, the Lorenz curves deviate from the diagonal line of equal participation, indicating that bounty creation is not evenly distributed across supporters.
The curves are close to one another, but they suggest modest differences in concentration across categories.
In particular, the Lorenz curve for deepfake bounties lies modestly above those for SFW and NSFW bounties, indicating that deepfake requests are somewhat more concentrated among a smaller set of highly active supporters.

This pattern is also visible at the 20th supporter percentile: approximately 56\% of deepfake bounties are posted by the top 20\% of supporters, compared to roughly 53\% for SFW bounties and 49\% for NSFW bounties.
The Gini coefficients provide a complementary summary of these distributions, with deepfake bounties showing the highest concentration ($G = 0.45$), followed by SFW ($G = 0.41$) and NSFW ($G = 0.37$).
Because these categories are not mutually exclusive (see Section~\ref{sec:methods}), Figure~\ref{fig:fig5} reflects concentration patterns across content categories rather than independent user populations.

\subsection{Deepfake intervention effectiveness}

We next examine how Civitai's platform-provided intervention is applied to deepfake bounty requests (\textbf{RQ5}).
As described in Section~\ref{sec:methods}, Civitai displays an informational alert on some bounties that request the reproduction of a real person's likeness.

\begin{figure}
\centering
\includegraphics[width=.7\linewidth]{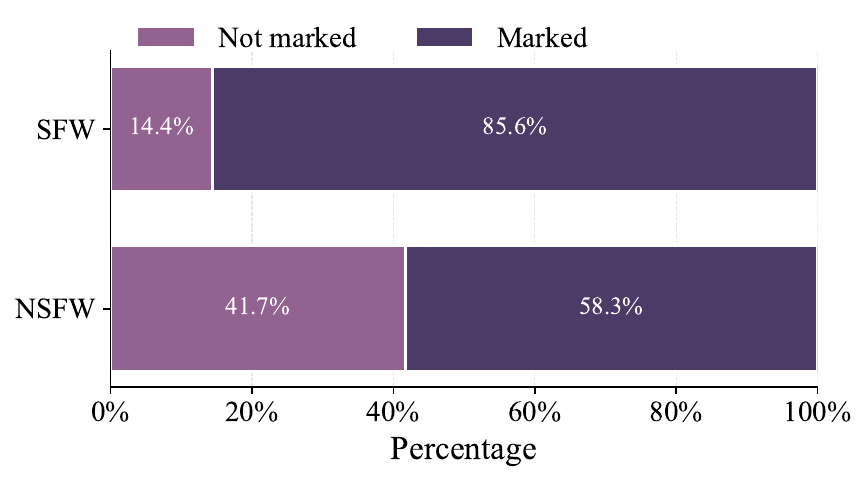}
\caption{Proportions of SFW and NSFW deepfake bounties that display the platform's informational alert. Civitai's deepfake intervention is absent from a substantial fraction of deepfake bounty requests, particularly for NSFW content. 
}
\label{fig:fig3}
\end{figure}

Figure~\ref{fig:fig3} shows the prevalence of this alert across SFW and NSFW deepfake bounties.
Among SFW deepfake requests, the alert is present on the large majority of bounties (85.6\%).
In contrast, only 58.3\% of NSFW deepfake bounties display the alert, leaving a substantial share (41.7\%) of NSFW deepfake requests unmarked.

These results reveal a pronounced disparity in the application of Civitai's deepfake intervention, with a large fraction of NSFW deepfake bounties lacking the platform-displayed alert.

\section{Discussions}
\label{sec:disc}

By analyzing Civitai bounties, this study provides the first systematic and quantitative look at a unique marketplace that coordinates demand, labor, and incentives for AI-generated content, including adult material and deepfakes. 
Unlike prior work that focuses on generated outputs or models shared on the platform, our analysis centers on demand-side signals (what users explicitly request) and platform permissiveness, offering insight into how content production is shaped by platform design and incentive structures. 

\subsection{LoRAs as tools for steering generative behavior}

We find that the majority of bounties request Low-Rank Adaptations (LoRAs), underscoring their central role in the Civitai ecosystem.
LoRAs are particularly consequential because they allow users to steer generative models toward behaviors they were not explicitly designed or intended to perform.
This includes generating images of specific real individuals, producing mature or explicit content, and exerting fine-grained control over attributes such as body positions or sexualized contexts.
In this sense, LoRAs function not merely as lightweight extensions of base models, but as mechanisms for bypassing or reshaping model guardrails through fine-tuning rather than prompting alone.

The prominence of LoRA-focused bounties suggests that users are not only seeking outputs, but also reusable tools that encode these behaviors and can be shared or redeployed beyond a single request.
This dynamic has important implications for governance: once created, such artifacts can persist, circulate, and enable repeated content generation well beyond the original bounty transaction. 
Prior work~\cite{hawkins2025deepfakes} emphasizes that LoRAs are often used to create deepfake variants contributing to a large portion of deepfakes in the wild.

\subsection{Prevalence and growth of NSFW content}

Across bounty types, adult (NSFW) requests constitute a substantial portion of the marketplace---nearly half of our entire dataset and a majority of video bounties in particular.
More notably, the proportion of NSFW bounties has been increasing steadily over time, now representing a majority of weekly requests.
This trend suggests that NSFW use is not a marginal or transient phenomenon, but an increasingly central use case within the bounty system.

While only a small fraction of bounties explicitly request adult deepfakes, the broader growth of NSFW requests raises concerns about how incentive-based marketplaces may amplify demand for increasingly explicit content over time. 
Future work could examine whether such growth reflects shifting community norms, increased technical capacity, or strategic behavior in response to platform moderation practices.

\subsection{Gendered targeting and ethical concerns around deepfakes}

Public figures, particularly actors and actresses, face heightened vulnerability to deepfake exploitation, likely due to the wide availability of their images online. 
Our analysis reveals a striking gender imbalance: deepfake targets are overwhelmingly female, with an approximate 9:1 ratio of women to men.
Both of these findings are consistent with prior research identifying nearly 35 thousand publicly downloadable deepfake model variants---96\% targeting women---on Civitai, collectively contributing close to 15 million downloads since November 2022~\cite{hawkins2025deepfakes}. 

These patterns align with broader evidence that deepfake technologies disproportionately harm women and are frequently used in non-consensual or sexually exploitative contexts~\cite{laffier2023deepfakes, kira2024deepfakes, lazard2025deepfake, chapman2024unveiling}. 
Even when requests are classified as SFW, they may still involve the generation of images without the knowledge or consent of the targeted individual.

An additional set of cases highlights how users themselves attempt to navigate the ethical boundaries of deepfake requests---often unsuccessfully.
We identify four bounties in which the targeted individual was described as the requester themselves or their spouse.
In three of these cases, supporters claimed to be requesting deepfake generation of their own likeness (two SFW and one implicitly open to NSFW depictions, based on discussion context).
In the fourth case, after users publicly questioned the request's ethics in the discussion section, the bounty supporter asserted that the example images depicted their wife and that the request was therefore acceptable.

These cases expose unresolved tensions within the platform's governance model.
They suggest that users are aware of the ethical stakes and may be attempting to justify their requests in the absence of clear procedural safeguards.
At the same time, such claims are difficult to verify, may be misleading, and---even if truthful---do not provide any mechanism for the purportedly targeted individuals to independently confirm consent or request removal. 
Responsibility for ethical judgment is thus implicitly shifted onto users, who may lack both the tools and authority to meaningfully address consent, deception, and harm within a public, incentive-driven marketplace.

\subsection{Concentration of participation and behavioral risk}

Our analysis also reveals that bounty creation is not evenly distributed across supporters.
Examining SFW, NSFW, and deepfake bounties, a minority of users account for a disproportionate share of requests. 
At the same time, modest differences appear across categories, with deepfakes displaying the highest concentration relative to SFW and NSFW categories.

From a governance perspective, these patterns motivate future work on whether behavioral signals, such as frequent requests, can help platforms prioritize review or intervention efforts, particularly for high-stakes content categories.
Such approaches would complement content-based moderation by incorporating information about user activity patterns over time.

\subsection{Platform interventions and enforcement gaps}
\label{sec:4.5}

Civitai has implemented an intervention intended to moderate deepfake-related requests by displaying an informational notice on bounties that involve reproducing a real person's likeness.
While this mechanism exists, its application is inconsistent.
We find that the notice is absent from almost half of NSFW deepfake bounties, indicating substantial gaps in how the intervention is applied.

These gaps point not to an absence of policy, but to limitations in enforcement.
During the study period (between 17--22 August 2024), Civitai formally prohibited bounties requesting depictions of real individuals,\footnote{\href{https://web.archive.org/web/20240822100938/https://civitai.com/content/tos}{web.archive.org/web/20240822100938/https://civitai.com/content/tos}} yet such requests persisted.
Both SFW and NSFW deepfake bounties are still present on the platform at the time of writing. 
While we cannot observe Civitai's internal moderation processes due to the scope of our data, this highlights the limits of policy updates without consistent enforcement. 
Our framework will make it easy to repeat the present analysis in the future, and our present results can serve as a baseline for evaluating future policy changes. 

More broadly, our findings underscore the challenges of moderating request-based systems, where supporters may strategically frame or obscure requests and responders may submit NSFW content even when it is not explicitly solicited.
Effective governance likely requires combining content-based rules with behavioral signals and proactive monitoring.
Future research is needed to develop methods for detecting subtle or non-obvious forms of deepfake policy violations that may evade existing safeguards.

\subsection{Limitations and future directions}
\label{sec:future}

Our analysis has several limitations.
First, our design focuses on what is requested and does not allow us to explore personal motivations or intent, nor final output usage. 
Second, collection ended on January 30, 2025 so that we could start our analysis.
Therefore, our data is not a complete representation of the current bounty space. 
Civitai extended their stricter deepfake policy (cf.~Section~\ref{sec:4.5}) from bounties to the full platform in May 2025.\footnote{\href{https://civitai.com/articles/22350}{civitai.com/articles/22350}}
Future research should perform longitudinal analysis to illuminate policy-enforcement gaps and reveal whether external pressure yields substantive behavioral change or merely symbolic policy updates.
Third, a single human annotator was used both to select the model for assigning bounty themes and to validate the deepfake labels. 
This choice balances reliability with ethical considerations around exposure to graphic content as discussed in Section~\ref{sec:methods}, but could introduce some bias in the ground truth compared to having multiple annotators. 
Fourth, as discussed in Section~\ref{sec:labeling_themes}, we assign one label for each bounty based on a prioritization rule. 
Therefore, our method does not capture the fact that some bounties might span across multiple classes.
Finally, some bounty requesters discuss paying multiple entrants in the comments. 
These additional transactions are not captured in our collected data, therefore our payment analysis in Section~\ref{sec:results} gives only a lower-bound estimate of the incentives. 

Future work could examine how participation concentration evolves over time, how posters and responders are connecting with each other, whether high-activity users adapt to moderation strategies, and how platforms might deploy dynamic, behavior-aware interventions.

Additional work is also needed to explore thematic structure within the broader NSFW bounty space beyond deepfakes.  
By grounding these questions in empirical evidence about demand, participation, and enforcement, this study contributes to a deeper understanding of how generative AI platforms shape and are shaped by the incentives they create.



\section{Ethical considerations statement}

This research is exempt from IRB review (Indiana University protocol 24578). 
Our data collection complies with the robots.txt directives provided by the Civitai platform. 
To protect the rights, privacy, and well-being of any real human, the manuscript does not disclose any personally identifiable information. 
The dataset released with this study includes only the data that is necessary to reproduce our results by generating the manuscript figures. It does not include any personally identifiable information. 

To limit exposure to disturbing material, we restricted manual review of graphic content to a single annotator rather than requiring independent replication, accepting a tradeoff between inter-rater reliability and researcher well-being. The annotator was briefed in advance and could pause or discontinue at any point. Future work on similar platforms should establish formal researcher well-being protocols, including access to psychological support. 

For annotation purposes, we submitted data to OpenAI models via their API. 
Although all data submitted to OpenAI was publicly available on the web, research involving sexual content warrants careful ethical consideration. 
We note that 43\% of bounties were labeled as NSFW content by the Civitai platform. We used this metadata to label these bounties as NSFW. We provided bounty titles and descriptions to OpenAI to understand what fraction of the platform-labeled NSFW bounties contain deepfakes. However, the theme-labeling procedure did not involve sending any image URLs about these bounties to OpenAI. 
For bounties not labeled as NSFW content by Civitai, we submitted image URLs to OpenAI. This data may involve real individuals depicted without their consent. We acknowledge this as a residual concern and encourage future work to consider local, privacy-preserving annotation models when processing content involving identifiable individuals. 
For the purpose of checking the strictness of the labels given by the Civitai platform, we passed URLs of all unblurred images, including NSFW ones, to the OpenAI Content Moderation API along with titles and descriptions. 
Under the OpenAI policy at the time of analysis, data submitted via the API was not used to train foundation models; by default, the standard OpenAI API retains inputs and outputs for up to 30 days for abuse monitoring.\footnote{\href{https://developers.openai.com/api/docs/guides/your-data}{developers.openai.com/api/docs/guides/your-data}} 
While this policy precluded use of submitted data for model training, it does not guarantee that input-output pairs were not retained for other purposes. 

In addition to the above considerations, publishing our results creates a risk of informing bad actors about potentially harmful uses of platforms like Civitai, which could lead to further harm. On the other hand, this work informs the general public, including  potential victims and policymakers, about such harmful uses. In this trade-off, we believe that the benefits of alerting the public outweigh the potential risks.

\section{Generative AI usage statement}

We used GPT-4o and GPT-4.1 as part of our annotation pipeline, as discussed in Section~\ref{sec:methods}. 
No AI assistance was used in any other aspects of the research or writing. 

\section{Acknowledgments}

We are grateful to Jisun An, Haewoon Kwak, and Kristina Lerman for helpful discussion. 
This work was supported in part by the Knight Foundation.

\section{Author contributions}

MRD led the conceptualization of the study, curated the dataset, and supervised the project. The final study design was developed collectively by MRD and SG. SG manually annotated the sample data, conducted the formal analysis and validation, developed visualizations with input from MRD, and drafted the manuscript. FM secured funding and oversaw project administration. MRD and FM reviewed and edited the manuscript.

\section{Code and data availability}

Our code and data are publicly available at \href{https://github.com/osome-iu/civitai_bounties_osome}{github.com/osome-iu/civitai\_bounties\_osome}.

\bibliographystyle{abbrvnat}
\bibliography{main.bib}

\clearpage

\appendix

\section*{Appendix}
\section{Model-human agreement on bounty themes}
\label{app:f1}

We report overall macro-F1 and by-label F1 scores comparing each model to the human annotator. 
We also report the weighted-F1 as there is class imbalance in the annotated sample, as reported in Table~\ref{tab:model-performance}.
While we use GPT-4.1 for the large scale theme extraction based on its better performance, we manually review all the flagged `Human deepfake' bounties.
43\% of the bounties were already identified as `NSFW' using platform-provided metadata, hence they were not part of the GPT-4.1-based classification.

\begin{table}[ht]
\centering
\caption{Classification performance by theme. For each model (GPT-4o and GPT-4.1), we report by-label F1 for each theme along with macro-F1 and weighted-F1.}
\label{tab:model-performance}
\renewcommand{\arraystretch}{1.3}
\begin{tabular}{lccc}
\hline
\textbf{Category} & \textbf{Number of samples} & \textbf{By-label F1 (GPT-4o)} & \textbf{By-label F1 (GPT-4.1)}  \\
\hline
Human attributes         & 6 & 0.80 & 0.91\\
Human deepfake           & 12 & 0.91 & 1.00\\
NSFW                     & 31 & 0.64 & 0.58\\
Fictional characters     & 64 & 0.88 & 0.91\\
Scenes, objects, clothing & 14& 0.85 & 0.97\\
Style and culture        & 10 & 0.55 & 0.74 \\
Miscellaneous            & 1 & 0.00 & 0.00 \\
Open requests            & 12 & 0.59 & 0.87 \\
\hline\hline
Macro-F1                 & & 0.65 & 0.75 \\
Weighted-F1              & & 0.77 & 0.84 \\
\hline
\end{tabular}
\end{table}

\section{Prompts for general theme extraction}
\label{app:B}

\subsection*{System prompt}%
\begin{lstlisting}[basicstyle=\ttfamily\itshape\small, breaklines=true, frame=none, breakindent=0pt, aboveskip=0pt]
Civitai is an online community where users collaborate on generative AI projects. Bounties on Civitai are challenges that offer rewards for completing specific tasks. You are an expert classifier responsible for assigning thematic labels to bounties based on their TITLE, DESCRIPTION, and associated IMAGES. The TITLE and DESCRIPTION will be in HTML format; IMAGES will be provided as URL links.
\end{lstlisting}

\begin{lstlisting}[basicstyle=\ttfamily\itshape\small, breaklines=true, frame=none]
Themes:
- human_attributes: Features or characteristics related to human beings, such as facial expressions, body types, or human-like traits.
- human_deepfake: Realistic representations of real people or requests for deepfakes of specific real individuals.
- nsfw: Explicit, mature, or adult content that is not safe for work environments.
- fictional_characters: Non-human, fictional, cartoon, or virtual characters, including those from movies, games, books, or anime.
- scene_objects_clothing: Scenery, objects, clothing, accessories, or decorative items.
- style_and_culture: Art styles, artistic movements, or cultural elements.
- open_request: General requests without specific or clear instructions.
- miscellaneous: Content that does not fit any of the above categories.

Classification instructions:
1. Carefully analyze the TITLE, DESCRIPTION, and IMAGES.
2. Assign the single most appropriate theme from the list above.
3. If clearly applicable, you may assign up to two additional themes, each with a brief rationale.
4. Follow the prioritization rules below when selecting the best-fit theme:
   a. First, check if the bounty involves realistic representations of specific real individuals. If so, assign 'human_deepfake' as the best-fit theme.
   b. If not a deepfake, check whether NSFW content is implied, allowed, or requested. If so, assign 'nsfw' as the best-fit theme.
   c. If neither condition applies, choose the theme that best captures the primary visual or stylistic focus of the bounty.
5. If no theme clearly fits, assign 'miscellaneous'.
\end{lstlisting}

\subsection*{User prompt}%
\begin{lstlisting}[basicstyle=\ttfamily\itshape\small, breaklines=true, frame=none, aboveskip=0pt]
TITLE: {title}

DESCRIPTION: {description}
\end{lstlisting}

\section{Prompts for deepfake extraction}
\label{app:C}

\subsection*{System prompt}
\begin{lstlisting}[basicstyle=\ttfamily\itshape\small, breaklines=true, frame=none, aboveskip=0pt]
You are an expert classifier for Civitai 
bounties. Classify bounties based on 
TITLE and DESCRIPTION (may contain HTML).

## Task
Determine if the bounty requests content about a 
specific person or person-like entity 
(real people, named fictional characters, 
VTubers, mascots).

## Classification Logic
    
**If NO specific person/character is identified:**
Return:
{
    "Deepfake": false,
    "Name": null,
    "Occupation": null,
    "RealHuman": false,
    "Rationale": "Brief explanation why no 
    specific person identified <= 50 words)"
}

**If a specific person/character IS identified:**
    
- **Deepfake**: 
    - true ONLY if requesting 
    generation/imitation of a REAL human's 
    likeness/voice/body.
    - false for: fictional characters, "style of 
    X" (without replicating likeness), 
    generic celebrity references
    
- **Name**: Most specific full name or canonical 
character name (choose primary if multiple)
    
- **Occupation**: Short role (e.g., "actor", 
"politician", "cartoon character", "VTuber")
    
- **RealHuman**: 
    - true = real human individual
    - false = fictional/animated/virtual OR when 
    Deepfake is false

- **Rationale**: REQUIRED. Explain your 
classification in <= 50 words.

## Critical Rules
- ALWAYS include Rationale field
- If Deepfake=false, then RealHuman MUST be false
- Ignore HTML tags; use visible text only
- "Style of X" without likeness replication -> 
Deepfake=false
- Generic references (no named person) -> all 
fields false/null

## Output Format
Return ONLY valid JSON with lowercase booleans and these exact keys:
{
    "Deepfake": <bool>, 
    "Name": <str|null>, 
    "Occupation": <str|null>, 
    "RealHuman": <bool>, 
    "Rationale": "<string>"
}

## Examples
Input: "Create Scarlett Johansson portrait" / 
"Photoreal headshot of Scarlett Johansson"
Output: 
{
    "Deepfake": true, 
    "Name": "Scarlett Johansson", 
    "Occupation": "actor", 
    "RealHuman": true, 
    "Rationale": "Requests photorealistic 
    likeness of real actor Scarlett Johansson, 
    which qualifies as deepfake."
}

Input: "Anime Spider-Man" / "Cell-shaded 
Spider-Man action pose"
Output: 
{
    "Deepfake": false, 
    "Name": "Spider-Man", 
    "Occupation": "cartoon character", 
    "RealHuman": false, 
    "Rationale": "Spider-Man is fictional 
    character, not a real human, so not a 
    deepfake request."
}

Input: "Fitness model poses" / "Generic model 
reference, no specific person"
Output: 
{
    "Deepfake": false, 
    "Name": null, 
    "Occupation": null, 
    "RealHuman": false, 
    "Rationale": "No individual person 
    identified; only generic fitness model 
    references."
}

Input: "Portrait in the style of Rembrandt" / "Use Rembrandt's painting technique"
Output: 
{
    "Deepfake": false, 
    "Name": "Rembrandt", 
    "Occupation": "painter", 
    "RealHuman": false, 
    "Rationale": "Requests artistic style only, 
    not replicating Rembrandt's likeness or body."
}
\end{lstlisting}

\subsection*{User prompt}

\begin{lstlisting}[basicstyle=\ttfamily\itshape\small, breaklines=true, frame=none]
TITLE: {title}

DESCRIPTION: {description}
\end{lstlisting}

\end{document}